\documentclass[a4paper,fleqn,usenatbib]{mnras}
\usepackage{amsmath}
\usepackage{mathptmx}
\usepackage{txfonts}
\usepackage[T1]{fontenc}
\usepackage{ae,aecompl}
\usepackage{graphicx}	% Including figure files
\usepackage{amssymb}	% Extra maths symbols
\usepackage[normalem]{ulem}
\newcommand{\be}{\begin{equation}}
\newcommand{\e}{\end{equation}}
\newcommand{\bear}{\begin{eqnarray}}
\newcommand{\ear}{\end{eqnarray}}

\newcommand{\del}{\partial}

\def\aj{AJ}
\def\apj{ApJ}
\def\apjs{ApJS}

\def\mnras{MNRAS}
\def\nar{New Astronomy Reviews}
\def\aap{A\&A}
\def\prd{Physical Review D}

\def\apjs{ApJS}
\def\apjl{ApJ Letters}

\title{} \author{} \title[Mass density and $\Lambda$ from entropy] {A
  new method to probe the mass density and the cosmological constant
  using configuration entropy}
    
\author[Pandey, B. and Das, B.] {Biswajit Pandey
  \thanks{E-mail:biswap@visva-bharati.ac.in} and {Biswajit Das
    \thanks{E-mail:bishoophy@gmail.com}} \\ Department of Physics,
  Visva-Bharati University, Santiniketan, Birbhum, 731235, India\\ }
   
 \date{\today}

 \pubyear{2018}
  
\begin{document}
\label{firstpage}
\pagerange{\pageref{firstpage}--\pageref{lastpage}} 

\maketitle

\begin{abstract}
We study the evolution of the configuration entropy for different
combinations of $\Omega_{m0}$ and $\Omega_{\Lambda0}$ in the flat
$\Lambda$CDM universe and find that the cosmological constant plays a
decisive role in controlling the dissipation of the configuration
entropy. The configuration entropy dissipates at a slower rate in the
models with higher value of $\Omega_{\Lambda0}$. We find that the
entropy rate decays to reach a minimum and then increases with
time. The minimum entropy rate occurs at an earlier time for higher
value of $\Omega_{\Lambda0}$.  We identify a prominent peak in the
derivative of the entropy rate whose location closely coincides with
the scale factor corresponding to the transition from matter to
$\Lambda$ domination. We find that the peak location is insensitive to
the initial conditions and only depends on the values of $\Omega_{m0}$
and $\Omega_{\Lambda0}$. We propose that measuring the evolution of
the configuration entropy in the Universe and identifying the location
of the peak in its second derivative would provide a new and robust
method to probe the mass density and the cosmological constant.

\end{abstract}

 \section {Introduction}
 
 Understanding the dark matter and dark energy remain the most
 challenging problems in cosmology. Observations suggest that the
 baryons or the ordinary matter constitutes only $\sim 5\%$ of the
 Universe. It is believed that $\sim \frac{1}{3}$ of the Universe is
 made up of the gravitating mass out of which $\sim 85\%$ is in the
 form of a hypothetical unseen matter dubbed as the ``dark
 matter''. The remaining $\sim \frac{2}{3}$ of the Universe is
 accounted by some mysterious hypothetical component dubbed as the
 ``dark energy''. The dark energy is believed to be responsible for
 driving the current accelerated expansion of the Universe. The
 existence and abundance of these mysterious components are determined
 from various observations.

 At present, the $\Lambda$CDM model where $\Lambda$ stands for the
 cosmological constant and CDM stands for the cold dark matter stands
 out as the most successful model in explaining most of the
 cosmological observations till date. The CDM model was initially
 introduced by \citet{peebles}. \citet{davis} carried out the
 pioneering numerical study of the CDM distribution which paved a new
 era allowing comparison of theory with multitude of observations.

The current paradigm of structure formation is supported by many
complementary observations. The fact that the CMBR angular power
spectrum peaks at $l \sim 200$ suggests a spatially flat Universe
where the mean energy density of the Universe must be close to the
critical density \citep{komatsu,planck}.  Other observations from
dynamics of galaxies and clusters \citep{carlberg}, the X-ray
observations of galaxy clusters \citep{mohr}, Sunyaev-Zeldovich effect
\citep{grego}, weak lensing \citep{benjamin, fu}, baryonic acoustic
oscillations (BAO) \citep{eisenstein1}, correlation functions
\citep{hawkins} and the power spectrum of density fluctuations
\citep{tegmark, reid, percival} revealed that the total mass density
parameter including both the baryonic and non-baryonic components must
be $\sim 0.3$. A flat Universe with $\Omega_m=0.3$ leaves us with no
other choice but $\Omega_{\Lambda}=0.7$ which fits the bill
perfectly. Further, the existence of dark energy is also supported by
independent observations such as Type Ia supernova \citep{riess,
  perlmutter} and BAO \citep{wang, eisenstein2} with very high
confidence.

The information entropy can be an useful tool for characterizing the
inhomogeneities in the mass distribution
\citep{hosoya,pandey2}. Recently, \citet{pandey1} propose that the
evolution of configuration entropy of the mass distribution in the
Universe may drive the cosmic acceleration. It has been argued that
the configuration entropy of the Universe decreases with time due to
the amplification of the density perturbations by the process of
gravitational instability. The configuration entropy continues to
dissipate in a matter dominated Universe. The dissipation of
configuration entropy due to this transition from smoother to clumpier
state demands existence of some efficient entropy generation
mechanisms to counterbalance this loss. If the other entropy
generation mechanisms are not sufficient to counter this loss then the
Universe must expand in such a way so as to prevent the further growth
of structures and stop the leakage of information
entropy. Interestingly, the dissipation of the configuration entropy
comes to a halt in a $\Lambda$ dominated Universe due to the
suppression of the growth of structures on large scales. Recently
\citet{das} used the evolution of the configuration entropy to
distinguish various dynamical dark energy parameterizations. The
importance and some interesting implications of inhomogeneities in
cosmology has been highlighted earlier in \citet{buchert1} and
\citet{buchert2}.

In the present work, we assume that the $\Lambda$CDM model with flat
FRW metric to be the correct model of the Universe and the density
parameters associated with matter and $\Lambda$ are to be determined
from observations. We propose a new method for the determination of
the density parameter associated with the mass and the cosmological
constant. The method is based on the study of the evolution of the
configuration entropy in the Universe. In future, the present
generation surveys like SDSS \citep{york}, 2dFGRS \citep{colles}, dark
energy survey \citep{abbott} combined with various other future
surveys like DESI, Euclid and different future 21 cm experiments like
SKA would allow us to measure the configuration entropy at different
epochs and study its evolution. The method presented in this work
would provide an alternative route to measure the mass density and the
cosmological constant in an independent and unique way and compare
their values obtained by the other methods from various observations.

\section {Theory}
\subsection {Configuration entropy and its evolution}
The observations of the cosmic microwave background radiation (CMBR)
suggest that the Universe was highly uniform in the past. But the
matter distribution in the present day Universe is highly clumpy due
to the structure formation by gravity. \citet{pandey1} defines the
configuration entropy of the mass distribution following the idea of
information entropy \citep{shannon48} as,
 \begin {eqnarray}
   S_c(t) = - \int \rho(\vec{x},t)\log \rho(\vec{x},t)\, dV.
   \label{eq:one}
  \end {eqnarray}
Here $S_c(t)$ is the configuration entropy of the mass distribution at
time $t$ over a sufficiently large comoving volume $V$. The volume $V$
is divided into a large number of subvolumes $dV$ and the density
$\rho(\vec{x},t)$ is measured inside each volume element.

The distribution is treated as a fluid on large scales. The continuity
equation for the fluid in an expanding universe is given by,
  \begin {eqnarray}
   \frac {\partial \rho}{\partial t} + 3 \frac {\dot a}{a}\rho + \frac
         {1}{a}\nabla \cdot (\rho \vec {v}) = 0.
   \label{eq:two}
  \end {eqnarray}
 Here $a$ is the scale factor and $\vec {v}$ denotes the peculiar
 velocity of the fluid inside the volume element $dV$.
 
  If we multiply \autoref{eq:two} by $(1 + \log \rho)$ and integrate
  over the entire volume $V$, then we get the entropy evolution
  equation \citep{pandey1} as,
  \begin {eqnarray}
   \frac {dS_c(t)}{dt} + 3\frac{\dot a}{a}S_c(t) - \frac{1}{a}\int \rho (3 \dot a + \nabla \cdot \vec {v})\, dV = 0.
   \label{eq:three}
  \end {eqnarray}
 
  Changing variable from $t$ to $a$ in \autoref{eq:three} we get,
  \begin {eqnarray}
   \frac {dS_c(a)}{da}\dot a + 3\frac {\dot a}{a}S_c(a) - F(a) = 0.
   \label{eq:four}
  \end {eqnarray}
where $F(a)$ is given by,
\begin{eqnarray}
   F(a) = 3MH(a) + \frac {1}{a} \int \rho (\vec {x}, a)\nabla \cdot \vec {v}\, dV.
   \label {eq:five}
  \end{eqnarray}
 
  Here $M = \int \rho (\vec {x}, a)\, dV = \int \bar \rho (1 + \delta
  (\vec {x}, a))\, dV$ gives the total mass inside $V$ and $\delta
  (\vec {x}, a) = \frac {\rho (\vec {x}, a) - \bar \rho}{\bar \rho}$
  gives the density contrast in a subvolume $dV$ centred at the
  comoving co-ordinate $\vec{x}$. The $\bar \rho$ is the mean density
  of matter inside the comoving volume $V$.

 One can simplify \autoref{eq:four} further to get,
\begin {eqnarray}
  \frac {dS_c(a)}{da} + \frac {3}{a}(S_c(a) - M) + \bar\rho f(a) \frac
        {D^2(a)}{a}\int \delta^2(\vec x)\, dV = 0.
  \label {eq:six}
\end {eqnarray}
where, $D(a)$ is the growing mode of density perturbation and $f(a)=
\frac {dlnD}{dlna} = \frac {a}{D} \frac {dD}{da}$ is the dimensionless
linear growth rate.

We need to solve \autoref{eq:six} to study the evolution of the
configuration entropy for any given cosmological model. The
time-independent quantities in the third term of \autoref{eq:six} are
set to $1$ for the sake of simplicity. We calculate $D(a)$ and $f(a)$
for the cosmological model under consideration. We then numerically
solve the \autoref{eq:six} using the $4^{th}$ order Runge-Kutta
method.

The second and third term in \autoref{eq:six} together decides the
evolution of the configuration entropy. The second term is decided by
the initial condition whereas the third term is governed by the nature
of the growth of structures in a particular cosmological model. At the
initial stage, the second term solely dictates the evolution because
the growth factor remains negligible. The third term only comes into
play when the growth of structures becomes significant. So the
cosmology dependence of the configuration entropy arises purely from
the third term in \autoref{eq:six}.

The \autoref{eq:six} can be solved analytically ignoring the third
term which is given by,

\begin {eqnarray}
  \frac {S_c(a)}{S_c(a_i)} = \frac{M}{S_c(a_i)}+\Bigg(1-\frac{M}{S_c(a_i)}\Bigg)\Bigg(\frac{a_i}{a}\Bigg)^3.
  \label {eq:seven}
\end {eqnarray}
where, $a_i$ is the initial scale factor and $S_c(a_i)$ is the initial
entropy. We choose $a_i$ to be $10^{-3}$ throughout the present
analysis. According to this solution, we expect a sudden growth or
decay in the configuration entropy near the initial scale factor $a_i$
when $S_c(a_i)<M$ and $S_c(a_i)>M$ respectively. On the other hand, no
such transients are expected when $S_c(a_i)=M$. Since we are only
interested in the cosmology dependence of the configuration entropy,
we shall focus on the solution of \autoref{eq:six} for $S_c(a_i)=M$ in
the present work. The solutions in the other two cases are similar
other than the transients present near the initial scale factor.

\begin{figure*}
   \resizebox{8.5 cm}{!}{\rotatebox{0}{\includegraphics{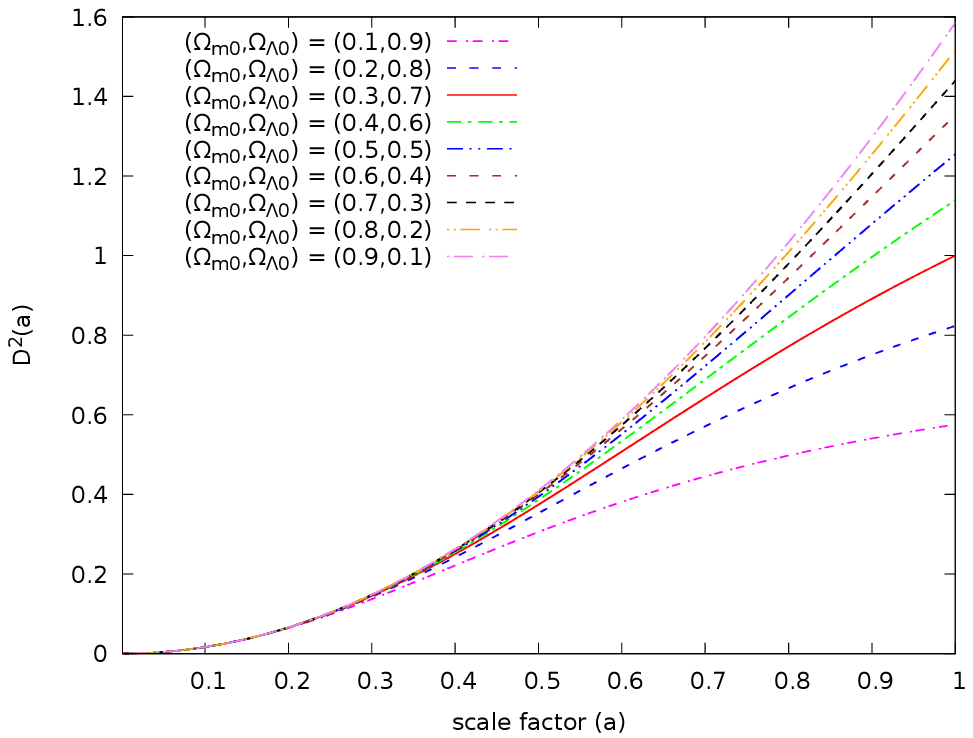}}}
   \hspace{0.5 cm}
   \resizebox{8.5 cm}{!}{\rotatebox{0}{\includegraphics{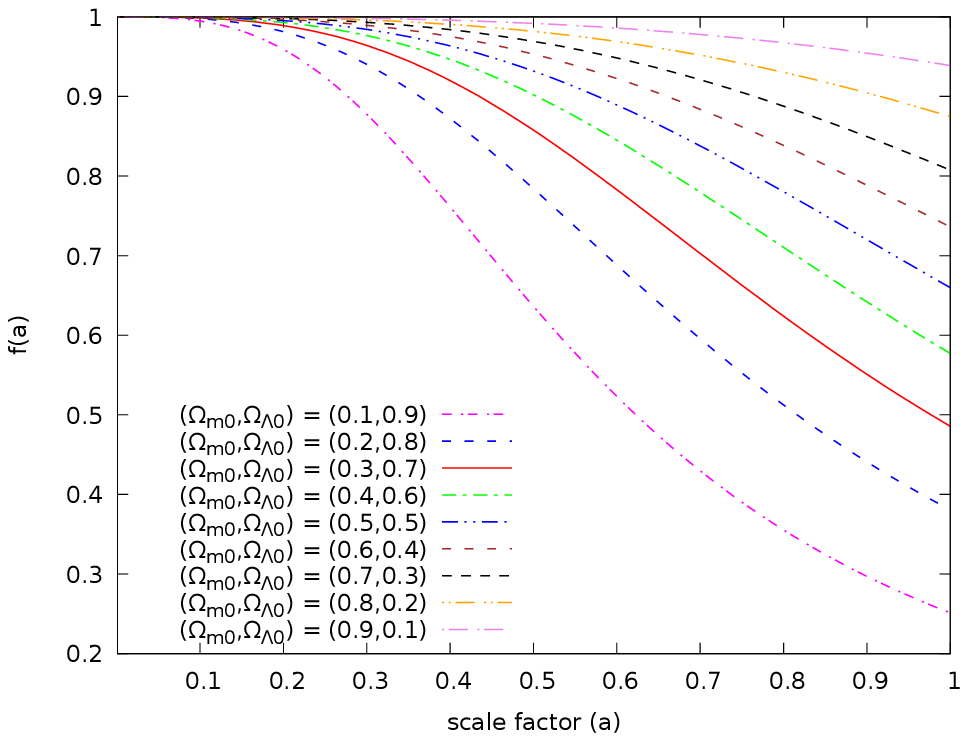}}}\\
   \caption{The left panel and the right panel of this figure
     respectively show $D^{2}(a)$ and $f(a)$ as a function of scale
     factor for different combinations of
     $(\Omega_{m0},\Omega_{\Lambda0})$ within the flat $\Lambda$CDM
     cosmology. The value of $D(a)$ is normalized to $1$ at present
     for the combination
     $(\Omega_{m0},\Omega_{\Lambda0})=(0.3,0.7)$. The $D(a)$ values in
     the other models are normalized with respect to this model.}
   \label{fig:d2f}
  \end{figure*}

 \begin{figure*}
   \resizebox{8.5 cm}{!}{\rotatebox{0}{\includegraphics{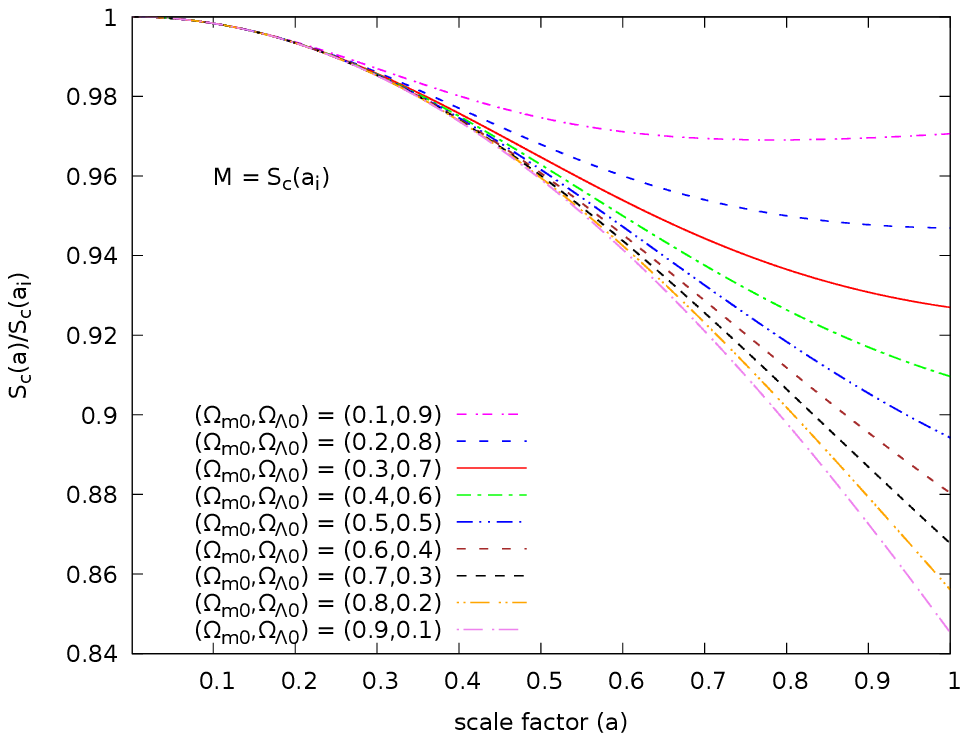}}}
   \hspace{0.5 cm}
   \resizebox{8.5 cm}{!}{\rotatebox{0}{\includegraphics{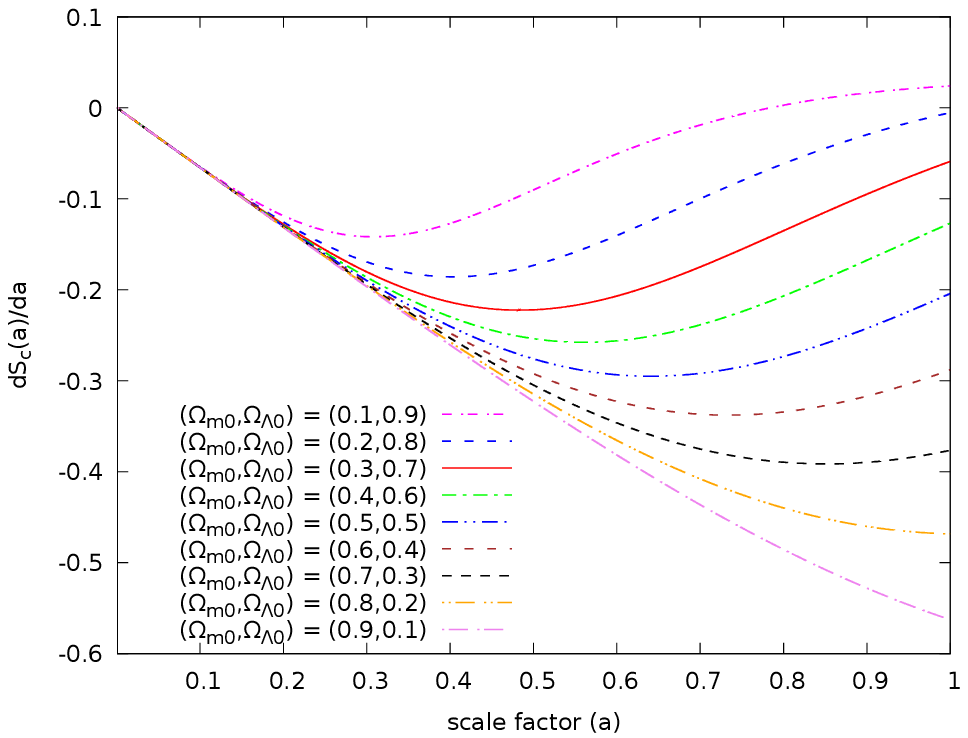}}}\\
   \vspace{-0.1 cm} 
   \resizebox{8.5 cm}{!}{\rotatebox{0}{\includegraphics{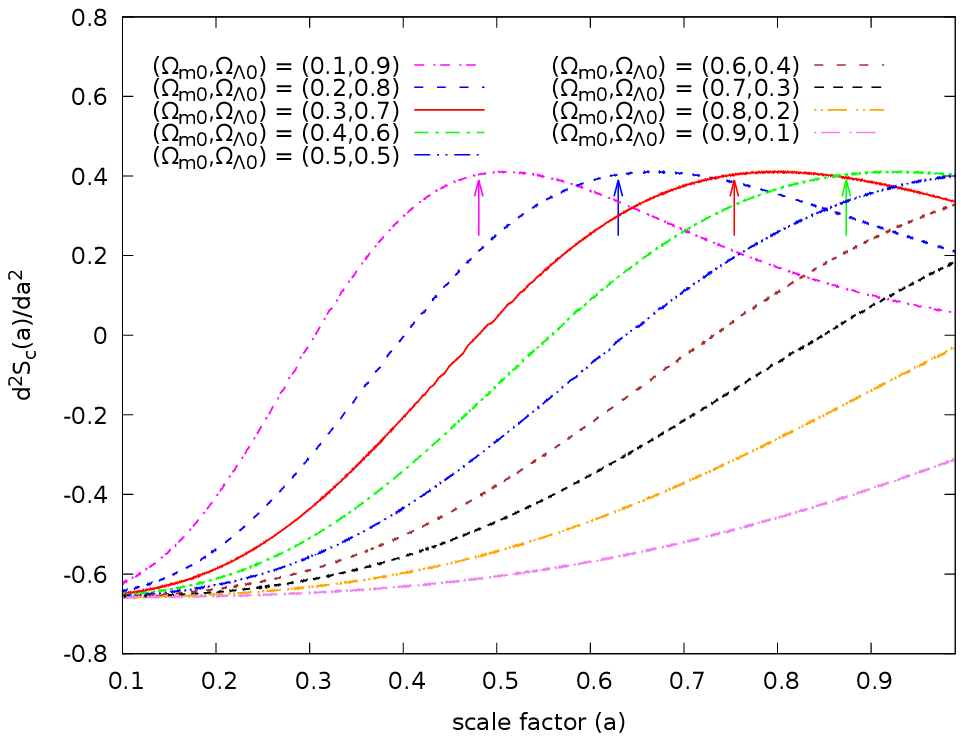}}}\\
   \caption{The top left panel of this figure shows the evolution of
     the configuration entropy for different combinations of
     $(\Omega_{m0},\Omega_{\Lambda0})$ within the flat $\Lambda$CDM
     model. The top right panel and the bottom middle panel
     respectively show the first and the second derivative of the
     configuration entropy in these models. The arrows in different
     colours in the bottom middle panel mark the scale factors
     $a_t=\Big(\frac{\Omega_{m0}}{\Omega_{\Lambda0}}\Big)^{\frac{1}{3}}$
     corresponding to the transition from matter to $\Lambda$
     domination in the respective models. The transition scale factors
     closely coincide with the peak locations in each model.}
   \label{fig:ent}
  \end{figure*}

\subsection{The growing mode and the dimensionless linear growth rate}
The CMBR observations show that the Universe is highly isotropic. But
the same observations also reveal that there are small anisotropies of
the order of $10^{-5}$ imprinted in the CMBR temperature maps. These
tiny fluctuations are believed to be the precursor of the large scale
structures observed in the present day Universe. The primordial
density perturbations were amplified by the process of gravitational
instability for billions of years. The growth of the density
perturbations $\delta(\vec{x},t)$ can be described by the linear
perturbation theory when $\delta<<1$. Considering only
perturbations to the matter sector, the linearized equation for the
growth of the density perturbation is given by,
 \begin {eqnarray}
   \frac{\del^{2}\delta(\vec{x},t)}{\del t^{2}} + 2H \frac{\del
     \delta(\vec{x},t)}{\del t} - \frac{3}{2}\Omega_{m0}
   H_{0}^{2}\frac{1}{a^{3}}\delta(\vec{x},t) = 0.
    \label{eq:linperturb}
  \end {eqnarray}
Here $\Omega_{m0}$ and $H_{0}$ are the present values of the mass
density parameter and the Hubble parameter respectively. This equation
has two solutions, one which grows and another which decays away with
time. The growing mode solution amplifies the density perturbations at
the same rate at every location so that the density perturbation at
any location $\vec{x}$ can be expressed as,
$\delta(\vec{x},t)=D(t)\delta(\vec{x})$. Here $D(t)$ is the growing
mode and $\delta(\vec{x})$ is the initial density perturbation at the
location $\vec{x}$.

The growing mode solution of \autoref{eq:linperturb} can be expressed
\citep{peebles_book} as,
\begin {eqnarray}
   D(a) = \frac{5}{2}\Omega_{m0} X^{\frac{1}{2}}(a) \int_0^{a} \frac{da^{\prime}}{a^{\prime 3} X^{\frac{3}{2}}(a^{\prime})},
  \label{eq:gmode}
  \end {eqnarray}
where $X(a)=\frac{H(a)^{2}}{H_{0}^{2}}=[\Omega_{m0}
  a^{-3}+\Omega_{\Lambda0}]$ in an Universe with only matter and
cosmological constant.

In a flat Universe, the dimensionless linear growth rate $f(a)=\frac{d
  \ln \delta}{d \ln a}$ can be well approximated \citep{lahav} by,
\begin {eqnarray}
   f(a) = \Omega_m(a)^{0.6} + \frac{1}{70}[1-\frac{1}{2}\Omega_m(a)(1+\Omega_m(a))],
  \label{eq:grate}
  \end {eqnarray}
where the matter density history $\Omega_m(a)$ can be written as
$\Omega_m(a)=\frac{\Omega_{m0}a^{-3}}{X(a)}$.

In the present work, we consider different combinations of
$\Omega_{m0}$ and $\Omega_{\Lambda0}$ within the framework of the flat
$\Lambda$CDM model and calculate $D(a)$ and $f(a)$ in each case. The
$D^{2}(a)$ and $f(a)$ for different models are shown in
\autoref{fig:d2f}. These are then used to solve \autoref{eq:six} to
study the evolution of the configuration entropy in each model.

 \section{Results and Conclusions}
We show the evolution of $\frac{S_c(a)}{S_c(a_i)}$ with scale factor
in the top left panel of \autoref{fig:ent}. We see that the
configuration entropy initially decreases with time. The dissipation
of the configuration entropy is driven by the growth of structures.
The dissipation is higher in the models with a larger value of
$\Omega_{m0}$. This is directly related to a higher growth factor
$D(a)$ and growth rate $f(a)$ in the models with a larger
$\Omega_{m0}$ (\autoref{fig:d2f}).  Clearly, the dissipation is less
pronounced in the models with larger $\Omega_{\Lambda0}$. The
structure formation is the outcome of two competing effects: one is
the tendency of the overdense regions to collapse under their self
gravity and the other is the tendency to move apart with the
background expansion. The cosmological constant $\Lambda$ contributes
to the later and thus resists the leakage of the configuration entropy
in the Universe by increasing the Hubble drag and suppressing the
structure formation on large scales.

We show the rate of change of the configuration entropy in the top
right panel of \autoref{fig:ent}. We find that the cosmological
constant $\Lambda$ plays an influential role in controlling the
dissipation of the configuration entropy. Models with larger value of
$\Omega_{\Lambda0}$ and smaller value of $\Omega_{m0}$ show a dip in
the slope of the configuration entropy at a smaller value of the scale
factor. For example in the model with $\Omega_{\Lambda0}=0.9$,
initially the slope decreases with increasing scale factor reaching
the minimum at $a \sim 0.3$. The slope then turns upward remaining
negative upto $a\sim 0.8$ thereafter upcrossing the zero. It then
slowly plateaus towards a stable value. A similar trend is observed
for the other models with different combinations of $\Omega_{m0}$ and
$\Omega_{\Lambda0}$. The minimum occur at $a \sim 0.4$ and $a \sim
0.5$ in the models with $\Omega_{\Lambda0}=0.8$ and
$\Omega_{\Lambda0}=0.7$ respectively. The minimum of the slope
indicates the time since $\Lambda$ becomes proactive in suppressing
the dissipation of the configuration entropy. The minimum appears at a
smaller value of scale factor in the higher $\Omega_{\Lambda0}$ models
simply because the presence of the $\Lambda$ term is felt earlier in
these models.

The bottom middle panel of \autoref{fig:ent} shows the derivative of
the slopes shown in the top right panel of the figure. Initially, the
derivative of the slopes are negative for all the models which imply
that the slopes are decreasing with time. But the second derivative
keeps on increasing with time and eventually upcrosses zero at some
value of the scale factor. This scale factor corresponds to the
minimum of the slope. For example the model with
$\Omega_{\Lambda0}=0.9$ exhibit the zero upcrossing of the second
derivative at $a\sim 0.3$ where a minimum was observed in the
slope. Similarly, a zero upcrossing can be seen at $a \sim 0.4$ and $a
\sim 0.5$ in the $\Omega_{\Lambda0}=0.8$ and $\Omega_{\Lambda0}=0.7$
models respectively. The positive value of the second derivatives
after this scale factor suggest that the slopes are increasing with
time. However, the slopes themselves remain negative after the
occurrence of the minimum. This suggest that the configuration entropy
still continues to dissipate with time but with a gradually decreasing
rate. The second derivatives of the configuration entropy do not
increase monotonically but show a prominent peak at a specific value
of the scale factor. The peaks are clearly identified in the models
with $\Omega_{\Lambda0}=0.9$, $\Omega_{\Lambda0}=0.8$,
$\Omega_{\Lambda0}=0.7$ and $\Omega_{\Lambda0}=0.6$ at the scale
factor $a=0.48$, $a=0.63$, $a=0.75$ and $a=0.87$
respectively. Interestingly, the transition from matter to $\Lambda$
domination in the respective models are expected to occur at nearly
the same scale factors. It may be noted that the second derivatives
remain negative throughout the entire range of scale factor for
$\Omega_{\Lambda0}=0.1$ and $\Omega_{\Lambda0}=0.2$ models. The peaks
in these models and the rest of the models are expected to occur in
future and hence are not present in the figure. The dissipation rate
of the configuration entropy changes at a slower rate once the
$\Lambda$ domination takes place. This is related to the fact that the
growth of structures are completely shut off on larger scales once
$\Lambda$ begins to drive the accelerated expansion of the Universe.

We have repeated these analyses for $S_c(a_i)>M$ and $S_c(a_i)<M$ and
recovered the peaks at exactly the same locations. This indicates that
the peak locations are insensitive to the initial conditions and
depend only on the values of $\Omega_{m0}$ and $\Omega_{\Lambda0}$.

We expect that combining the measurements of the configuration entropy
at different redshifts from the present and future generation surveys
would enable us to study the evolution of the configuration entropy.
This would allow us to identify the location of the peak in its second
derivative and constrain the value of both the mass density and the
cosmological constant in the Universe.

We would like to point out here that the present method requires us to
measure the configuration entropy over a significantly large volume of
the Universe. This is to ensure that there are no net mass inflow or
outflow across the neighbouring volumes. The present studies suggest
that the Universe is homogeneous on large scales \citep{yadav, hogg,
  prakash, sarkar}. So at each redshift, a measurement of the
configuration entropy over a region extending few hundreds of Mpc
would be sufficient for the present analysis. However, the mutual
information between the spatially separated but causally connected
regions of the Universe may introduce a non-negligible dynamical
entanglement between them \citep{wiegand}. Any modification of the
evolution equation due to this entanglement need to be investigated
further.

Furthermore, the baryonic matter constitutes only a tiny fraction of the
matter budget. The baryonic matter distribution is expected to be
biased with respect to the dark matter distribution. Currently, the
proposed method requires us to map the distribution of an unbiased
tracer of the underlying mass distribution at multiple redshifts. The
introduction of bias may complicate the analysis which we would like
to address in a future work.

Finally, we conclude that the analysis presented in this work provides
an alternative avenue for the determination of the mass density and
the cosmological constant $\Lambda$ by studying the evolution of the
configuration entropy in the Universe.  We expect this to find many
useful applications in the study of the mysterious dark matter and the
elusive cosmological constant.

\section {Acknowledgement}
The authors thank an anonymous reviewer for valuable comments and
suggestions. The authors would like to acknowledge financial support
from the SERB, DST, Government of India through the project
EMR/2015/001037. BP would also like to acknowledge IUCAA, Pune and
CTS, IIT, Kharagpur for providing support through associateship and
visitors programme respectively.

% Don't change these lines
\bsp	% typesetting comment
\label{lastpage}

\begin{thebibliography}{99}
\bibitem[Abbott et al.(2018)]{abbott} Abbott, T.~M.~C., Abdalla,
  F.~B., Allam, S., et al.\ 2018, \apjs, 239, 18

\bibitem[Benjamin et al.(2007)]{benjamin} Benjamin, J., Heymans, C.,
  Semboloni, E., et al.\ 2007, \mnras, 381, 702

\bibitem[Buchert \& Ehlers(1997)]{buchert1} Buchert, T., \& Ehlers,
  J.\ 1997, \aap, 320, 1

\bibitem[Buchert(2000)]{buchert2} Buchert, T.\ 2000, General Relativity
  and Gravitation, 32, 105

\bibitem[Carlberg et al.(1996)]{carlberg} Carlberg, R.~G., Yee,
  H.~K.~C., Ellingson, E., et al.\ 1996, \apj, 462, 32

\bibitem[\protect\citeauthoryear{Colles et al.}{2001}]{colles} Colles,
  M. et al.(for 2dFGRS team) 2001,\mnras,328,1039

\bibitem[Das \& Pandey(2019)]{das} Das, B., \& Pandey, B.\ 2019,
  \mnras, 482, 3219

\bibitem[Davis et al.(1985)]{davis} Davis, M.,
  Efstathiou, G., Frenk, C.~S., \& White, S.~D.~M.\ 1985, \apj, 292,
  371

\bibitem[Eisenstein et al.(2005)]{eisenstein1} Eisenstein, D.~J.,
  Zehavi, I., Hogg, D.~W., et al.\ 2005, \apj, 633, 560

\bibitem[Eisenstein(2005)]{eisenstein2} Eisenstein, D.~J.\ 2005, \nar, 49, 360 

\bibitem[Fu et al.(2008)]{fu} Fu, L., Semboloni, E., Hoekstra, H., et
  al.\ 2008, \aap, 479, 9

\bibitem[Grego et al.(2000)]{grego} Grego, L., Carlstrom, J.~E., Joy,
  M.~K., et al.\ 2000, \apj, 539, 39

\bibitem[Hawkins et al.(2003)]{hawkins} Hawkins, E., Maddox, S., Cole,
  S., et al.\ 2003, \mnras, 346, 78

\bibitem[Hogg et al.(2005)]{hogg} Hogg, D.~W., Eisenstein, D.~J.,
  Blanton, M.~R., Bahcall, N.~A., Brinkmann, J., Gunn, J.~E., \&
  Schneider, D.~P.\ 2005, \apj, 624, 54

\bibitem[Hosoya et al.(2004)]{hosoya} Hosoya, A., Buchert, T., \&
  Morita, M.\ 2004, Physical Review Letters, 92, 141302

\bibitem[Komatsu et al.(2011)]{komatsu} Komatsu, E., Smith, K.~M.,
  Dunkley, J., et al.\ 2011, \apjs, 192, 18
  
\bibitem[Lahav et al.(1991)]{lahav} Lahav, O., Lilje, P.~B., Primack,
  J.~R., \& Rees, M.~J.\ 1991, \mnras, 251, 128


\bibitem[Mohr et al.(1999)]{mohr} Mohr, J.~J., Mathiesen, B., \&
  Evrard, A.~E.\ 1999, \apj, 517, 627

\bibitem[Pandey(2013)]{pandey2} Pandey, B.\ 2013, \mnras, 430, 3376 

\bibitem[Pandey(2017)]{pandey1} Pandey, B.\ 2017, \mnras, 471, L77

\bibitem[Peebles(1980)]{peebles_book} Peebles, P.~J.~E.\ 1980, The
  Large-scale Structure of the Universe, Princeton University Press,
  1980.~435 p.,

\bibitem[Peebles(1982)]{peebles} Peebles, P.J.~E.\ 1982, \apj, 263, L1

\bibitem[Percival et al.(2010)]{percival} Percival, W.~J., Reid,
  B.~A., Eisenstein, D.~J., et al.\ 2010, \mnras, 401, 2148

\bibitem[Perlmutter et al.(1999)]{perlmutter} Perlmutter, S.,
  Aldering, G., Goldhaber, G., et al.\ 1999, \apj, 517, 565

\bibitem[Planck Collaboration et al.(2016)]{planck} Planck
  Collaboration, Ade, P.~A.~R., Aghanim, N., et al.\ 2016, \aap, 594,
  A13

\bibitem[Reid et al.(2010)]{reid} Reid, B.~A., Percival, W.~J.,
  Eisenstein, D.~J., et al.\ 2010, \mnras, 404, 60

\bibitem[Riess et al.(1998)]{riess} Riess, A.~G.,
  Filippenko, A.~V., Challis, P., et al.\ 1998, \aj, 116, 1009

\bibitem[Sarkar et al.(2009)]{prakash} Sarkar, P., Yadav, J., 
Pandey, B., \& Bharadwaj, S.\ 2009, \mnras, 399, L128 

\bibitem[Sarkar \& Pandey(2016)]{sarkar} Sarkar, S., \& Pandey, B.\ 2016, \mnras, 463, L12 

\bibitem[Shannon(1948)]{shannon48} Shannon, C. E. \ 1948, Bell
System Technical Journal, 27, 379-423, 623-656

\bibitem[Tegmark et al.(2004)]{tegmark} Tegmark, M., Blanton, M.~R.,
  Strauss, M.~A., et al.\ 2004, \apj, 606, 702

\bibitem[Wang(2006)]{wang} Wang, Y.\ 2006, \apj, 647, 1 

\bibitem[Wiegand \& Buchert(2010)]{wiegand} Wiegand, A., \& Buchert,
  T.\ 2010, \prd, 82, 023523

\bibitem[Yadav et al.(2005)]{yadav} Yadav, J., Bharadwaj, S., Pandey,
  B., \& Seshadri, T.~R.\ 2005, \mnras, 364, 601

\bibitem[York et al.(2000)]{york} York, D.~G., et al.\ 2000, \aj,
  120, 1579

\end{thebibliography}
\end{document}